\documentclass[%
reprint,
showpacs,preprintnumbers,
showkeys,
amsmath,amssymb,
aps,
prl,
floatfix,superscriptaddress
]{revtex4-1}

\usepackage[english]{babel}
\usepackage[utf8]{inputenc}
\usepackage{graphicx}
\usepackage{hyperref}
\usepackage{xcolor}
\usepackage{amsmath}
\usepackage{amsfonts}
\usepackage{amssymb}
\usepackage{dsfont}
\usepackage{siunitx}
\usepackage{xcolor,import}
\usepackage{transparent}
\usepackage{todonotes}
\usepackage{sidecap} 
\usepackage{newtxtext,newtxmath}
\usepackage[T1]{fontenc}

\definecolor{refColor}{HTML}{EA00F2}
\definecolor{figColor}{HTML}{008DF2}
\definecolor{urlColor}{HTML}{00AEF2}

\hypersetup{
    unicode=false,                 
    pdftoolbar=true,               
    pdfmenubar=true,               
    pdffitwindow=false,            
    pdfstartview={FitH},           
    pdftitle={},                   
    pdfsubject={Quantum physics},  
    pdfkeywords={},                
    pdfnewwindow=true,             
    colorlinks=true,               
    linkcolor=figColor,            
    citecolor=refColor,            
    filecolor=magenta,             
    urlcolor=urlColor              
}

\newcommand{\bra}[1]{\left\langle #1\right|}             
\newcommand{\ket}[1]{\left| #1\right\rangle}              

\begin{document}

\title{Observation of collective decay dynamics of a single Rydberg superatom}

\author{Nina Stiesdal}
\affiliation{Department of Physics, Chemistry and Pharmacy, Physics@SDU, University of Southern Denmark, 5320 Odense, Denmark}
\author{Hannes Busche}
\affiliation{Department of Physics, Chemistry and Pharmacy, Physics@SDU, University of Southern Denmark, 5320 Odense, Denmark}
\author{Jan Kumlin}
\affiliation{Institute for Theoretical Physics III and Center for Integrated Quantum Science and Technology, University of Stuttgart, 70550 Stuttgart, Germany}
  \author{Kevin Kleinbeck}
\affiliation{Institute for Theoretical Physics III and Center for Integrated Quantum Science and Technology, University of Stuttgart, 70550 Stuttgart, Germany}
\author{Hans Peter B\"uchler}
\affiliation{Institute for Theoretical Physics III and Center for Integrated Quantum Science and Technology, University of Stuttgart, 70550 Stuttgart, Germany}
  \author{Sebastian Hofferberth}
\affiliation{Department of Physics, Chemistry and Pharmacy, Physics@SDU, University of Southern Denmark, 5320 Odense, Denmark}
\email{hofferberth@sdu.dk}
\date{\today}



\begin{abstract}
We experimentally investigate the collective decay of a single Rydberg superatom, formed by an ensemble of thousands of individual atoms supporting only a single excitation due to the Rydberg blockade. Instead of observing a constant decay rate determined by the collective coupling strength to the driving field, we show that the enhanced emission of the single stored photon into the forward direction of the coupled optical mode depends on the dynamics of the superatom before the decay. We find that the observed decay rates are reproduced by an expanded model of the superatom which includes coherent coupling between the collective bright state and subradiant states.
\end{abstract}

\maketitle

The collective interaction between an ensemble of emitters and photons is a fundamental topic of quantum optics, which has been extensively studied for over 50 years \cite{Dicke1954}. Collective enhancement of the emission, known as superradiance, has been observed in a variety of physical systems ranging from atoms \cite{Haroche1982} and ions \cite{Brewer1996} over molecules \cite{Sandoghdar2002}, artificial atoms coupled to microwave waveguides \cite{Wallraff2013}, and solid-state systems \cite{Hommel2007,Stobbe2016}, to ensembles of nuclei \cite{Roehlsberger2010}. Suppression of emission is more elusive because excitation fields typically do not couple to subradiant states, and was only recently observed for ensembles of more than two emitters \cite{Kaiser2016,Zheludev2017}.
Here, we investigate the collective emission of a single photon from a Rydberg superatom \cite{Kuzmich2012b} and show that the experimentally observed decay rate depends on the initial state-preparation by a few-photon driving field \cite{Hofferberth2017}. We attribute this effect to a coherent population redistribution between collective super- and subradiant states due to coherent excitation exchange between the individual emitters inside the superatom \cite{Adams2018,Buchler2018}.

A collectively excited ensemble features modifications to the rate and spatial distribution of its spontaneous emission \cite{Wilkowski2011,Wilkowski2014,Ye2016,Pillet2018}, and coherent exchange of photons between individual emitters in an ensemble results in a collective Lamb shift \cite{Lehmberg1970,Manassah1973,Scully2009,Roehlsberger2010,Ozeri2014,Havey2016,Adams2018b}. These phenomena can be understood in a semi-classical approach as dipole-dipole interaction between individual emitters \cite{Kivshar2013,Adams2015,Adams2016a,Yelin2017,Kimble2017,Kaiser2017,Adams2018}, or quantum-mechanically by treating the emitters as an interacting spin ensemble coupled to an optical mode \cite{Chang2015b,Pichler2015,Rey2016,Buchler2018}. The latter approach has been used to study the propagation of quantized light in one-dimensional waveguides, while the semi-classical approach enables investigation of large, weakly driven ensembles in two or three dimensions. In the single-excitation sector and as long as saturation of the medium can be neglected, the two approaches lead to equivalent results \cite{Scully2010,Kurizki2014}. There has recently been strong interest in structured emitter arrays for tailoring optical properties with unprecedented control \cite{Ruostekoski2012,Adams2016a,Yelin2017,Lukin2019}, for example exploiting subradiance to enhance photon storage fidelities \cite{Chang2017,Ruostekoski2016}. Striking experimental demonstrations of this concept are realisations of highly reflective monolayers with ultracold atoms in optical lattices \cite{Bloch2020} and solid state systems \cite{Imamoglu2018,Park2018}. 

\begin{figure}[ht]
\includegraphics{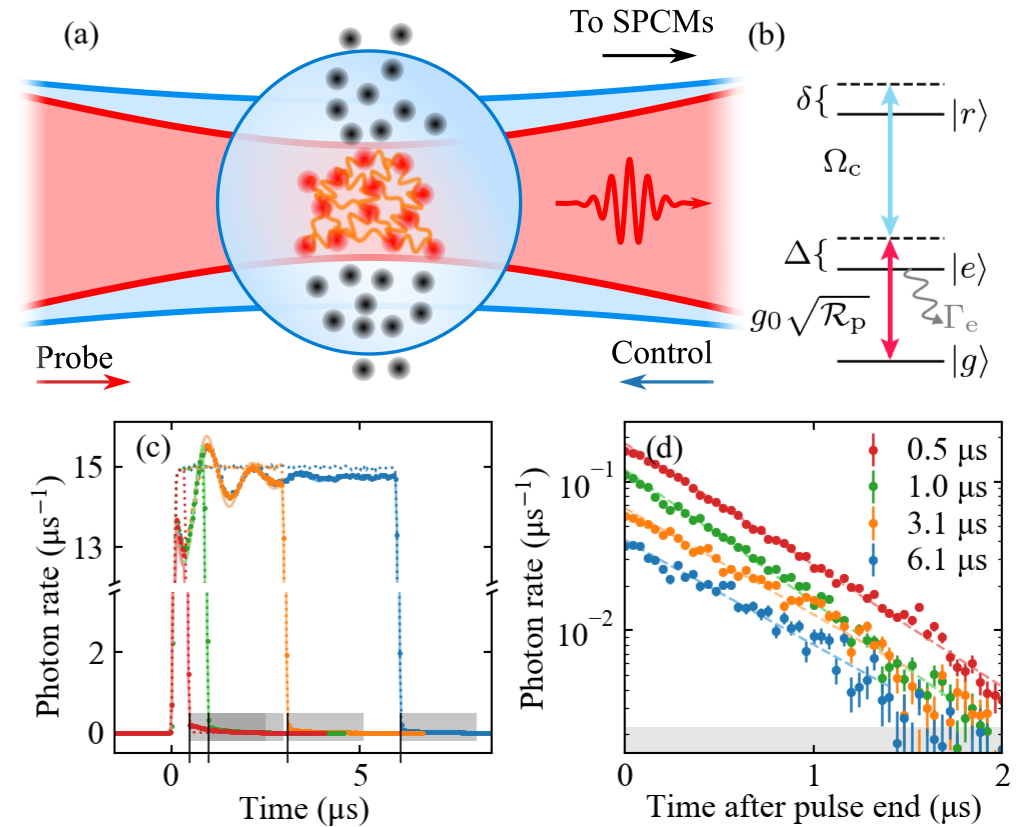}
\caption{\label{fig:Fig1} (a) Sketch of the experimental implementation and (b) single-atom level scheme to create a single Rydberg superatom. (c) Probe pulses of different duration measured on single-photon counters (SPCMs) with (solid circles) and without atoms (dashed lines). The solid lines show the solution to the master equation for the model system shown in Fig. \ref{fig:Fig2}. The shaded areas indicate the parts of the pulses our analysis focuses on. (d) Logarithmic plot of the difference between the signal with and without atoms detected after the driving pulse has ended for different pulse lengths. The time $t=0$ where the probe field is turned off, is extracted from fits to the probe pulse measured without atoms. The dashed lines are fits to the data points as described in the text. The grey area shows the level below which the data are excluded from the fits. The error-bars are standard error of the mean.}
\end{figure}

In this letter, we study collective directed single photon emission from an ultracold atomic ensemble which due to the Rydberg blockade effect \cite{Lukin2001} can only contain a single Rydberg excitation at any given time and is thus fully saturated by the absorption of a single photon \cite{Kuzmich2012b,Hofferberth2016e}. In this limit, the blockaded ensemble reduces to an effective two-level superatom with collectively enhanced coupling to the driving field \cite{Kuzmich2012c} and enhanced spontaneous decay into the forward direction of the driving mode. In contrast to the previous experimental study of the superatom driven by a few-photon probe \cite{Hofferberth2017}, we now study with very high accuracy the emitted light in the forward direction after the probe pulse. We experimentally observe that the rate of this collective emission depends on the duration and strength of the initial driving pulse, instead of being solely determined by the number of atoms $N$ forming the superatom. We attribute this effect to the redistribution of population between super- and subradiant collective states mediated by coherent photon exchange inside the superatom. To support this interpretation, we compare our data to an effective model that captures both the coherent internal dynamics as well as the overall dephasing of the collective excitation and find that this simple model captures the observed decay dynamics very well.

To create a single Rydberg superatom in our experiment, we prepare an optically trapped ensemble of $2 \times 10^4$ $^{87}$Rb atoms with dimensions of $\SI{6.5}{\micro m}$ along and $\SI{21}{\micro m} \times \SI{10}{\micro m}$ perpendicular to the probe direction ($1/\mathrm{e}$-radii of the Gaussian atomic density distribution), which is situated at the foci of counterpropagating probe ($1/\mathrm{e}^2$-waist radius $\SI{6.5}{\micro m}$) and strong control fields ($\SI{14}{\micro m}$) creating an excitation volume that is fully blockaded by a single Rydberg excitation, as shown in Fig. \ref{fig:Fig1} (a).
The probe and control fields, with variable probe photon rate $\mathcal{R_{\mathrm{p}}}$ and Rabi frequency $\Omega_\mathrm{c} = 2\pi \times \SI{13}{MHz}$ respectively, drive Raman transitions between the ground state $|g\rangle=|5S_{1/2},F=2,m_F=2\rangle$ and the Rydberg state $|r\rangle = |111S_{1/2},J=1/2,m_J=1/2\rangle$ via an intermediate state $|e\rangle = |5P_{3/2},F=3,m_F=3\rangle$, see Fig. \ref{fig:Fig1} (b). For large intermediate-state detuning $\Delta$, $\ket{e}$ can be adiabatically eliminated. Atoms in $|g\rangle$ are thus individually coupled to $|r\rangle$ with an effective Rabi frequency $\Omega = \sqrt{\mathcal{R_{\mathrm{p}}}}g_0\Omega_{\mathrm{c}}/(2\Delta)$ and Raman decay $\Gamma = \Gamma_{\mathrm{e}} (\Omega_{\mathrm{c}}/2\Delta)^2$ between $|r\rangle$ and $|g\rangle$. Here $g_0$ is the single-photon-single-atom coupling strength, and $\Gamma_{\mathrm{e}}= 2\pi \times \SI{6}{MHz}$ is the natural linewidth of $\ket{e}$. The transmission and re-emission of probe photons in the forward direction is detected using single photon counting modules (SPCMs) arranged in a Hanbury-Brown Twiss setup located behind a single mode fiber aligned onto the incoming probe mode.

The Rydberg blockade limits the ensemble to a single excitation and gives rise to its collective superatom nature with a collective ground state $\ket{G} = |g_1,\cdots, g_N\rangle$ and a singly excited collective bright state $\ket{W} = \tfrac{1}{\sqrt{N}}\sum^N_{j=1}\ket{g_1,\cdots,r_j,\cdots g_N}$, where $N \sim 5000$ is the number of atoms overlapping with the probe beam; see \cite{Hofferberth2017} for a more microscopic description of the bright state accounting for the atomic distribution and mode wave function. These collective states are coupled by the enhanced Rabi frequency $\Omega_{\mathrm{col}} = 2\sqrt{\kappa\mathcal{R_{\mathrm{p}}}}$ with collective single photon coupling $\sqrt{\kappa}=\sqrt{N} g_0\Omega_{\mathrm{c}}/(4\Delta)$.
This enhancement allows us to drive Rabi oscillations of the superatom between $\ket{G}$ and $\ket{W}$ with few-photon probe pulses \cite{Hofferberth2017}. The single-photon absorption and re-emission into the forward direction can be directly observed in the transmission of the Tukey-shaped probe pulses shown in Fig. \ref{fig:Fig1} (c), where $\mathcal{R_{\mathrm{p}}} = \SI{15.0}{\micro s^{-1}}$ and $\Delta = 2\pi \times \SI{100}{MHz}$. 

In the following, we investigate the collective decay dynamics after extinction of the probe pulse. Towards this end, we vary the duration of the Tukey-shaped probe pulse and record the forward-emitted light after extinguishing the probe light on a time scale shorter than the enhanced spontaneous decay rate $\kappa$ of the collective state, while the control field remains on. To confirm that the superatom at maximum emits one photon back into the probe mode, we analyse the photons statistics of the forward-emitted light after the driving pulse stops and obtain for the background-corrected second-order correlation function $g^{(2)}(\tau=0) < 0.1$ during the spontaneous decay of the superatom, consistent with single photon emission. We further confirm the presence of only a single excitation from the ion-counting statistics if the ensemble is field-ionized at the end of the probe pulse \cite{Hofferberth2016e}. In addition to collective emission, the superatom is also subject to non-radiative dephasing $\gamma_\mathrm{D}$ of the collectively enhanced state which results in non-retrievable storage of a single photon in the ensemble \cite{Hofferberth2016e}. 

Fig. \ref{fig:Fig1} (d) shows the probe light detected after extinction of the probe field for the four different pulse lengths shown in Fig. \ref{fig:Fig1} (c) (following subtraction of residual background light). As expected, we observe an exponential decay, with the initial amplitude determined by the bright state population. Additionally, though, we find that the rate of photon emission changes depending on the probe pulse duration and thus the internal state of the ensemble after the driving pulse. If the superatom's internal collective dynamics were fully incoherent, the decay rate would be constant and given by the sum of single photon coupling $\kappa$ and the rates for excited state population loss through Raman decay, $\Gamma$, and dephasing, $\gamma_{\mathrm{D}}$.

\begin{figure*}
\includegraphics{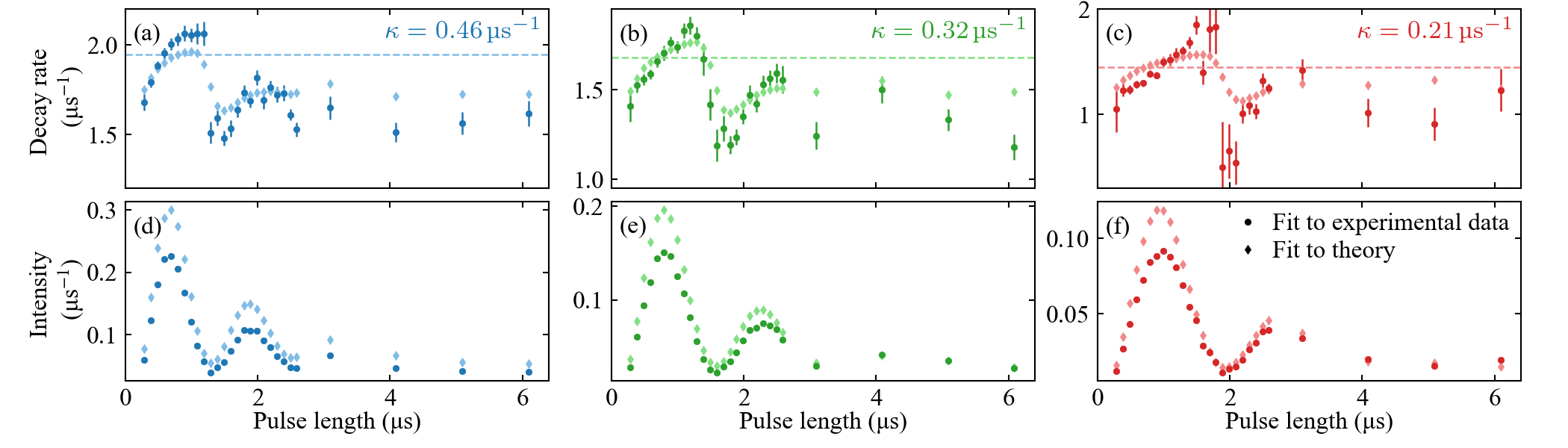}
\caption{\label{fig:Fig2} (a,b,c) Observed decay rates and (d,e,f) initial amplitudes as a function of pulse length extracted from exponential fits to data as shown in Fig. \ref{fig:Fig1} (c,d) for different values of the single-photon coupling strength $\kappa$. In addition to experimental data (dark circles), we show the theoretical results of the extented superatom model discussed in the text and shown in Fig. \ref{fig:Fig4} (light diamonds), as well as the constant decay rate of the simple superatom model without internal coherent dynamics (dashed lines). The errorbars shown on the rates are one standard deviation confidence interval of the exponential fits to the data.}
\end{figure*}

To systematically study the non-constant decay rate of the superatom, we repeat the above-described experiment with varying probe pulse length for a range of different parameter sets, varying the single-photon coupling strength $\kappa$ (Fig. \ref{fig:Fig2}) and the probe photon rate $\mathcal{R_{\mathrm{p}}}$ during the pulse (Fig. \ref{fig:Fig3}). To quantify the photon decay rate, we assume initially that the decay is exponential and fit the recorded photon rates with $I_0 \mathrm{e}^{-\gamma t}$ to extract the initial intensity $I_0$ (in photons/$\SI{}{\micro s}$) and the decay rate of forward emission $\gamma$. Because of the single-photon nature of the collective emission, we have to repeat the experiment for each parameter set sufficiently often to obtain mean photon traces as shown in Fig. \ref{fig:Fig1} (d). For all fits, we exclude data points below a threshold where the uncertainties become similar to the absolute values \cite{suppMat}, indicated by the shaded region in Fig. \ref{fig:Fig1}.

First, we show in Fig. \ref{fig:Fig2} the obtained amplitudes and rates as function of pulse length for three different values of the intermediate state detuning $\Delta = 2 \pi \times \SI{100}{MHz}$ (a, c), $\SI{125}{MHz}$ (b, e), and $\SI{150}{MHz}$ (c, f) and a fixed probe photon rate during the pulse of $\mathcal{R_{\mathrm{p}}}=\SI{15.0}{\micro s^{-1}}$. By changing $\Delta$, we vary both the coupling strength $\kappa$ between $\ket{G}$ and $\ket{W}$, as well as the Raman decay rate $\Gamma$. The initial amplitudes in Fig. \ref{fig:Fig2} (d, e, f) reflect the collective Rabi oscillation of the superatom during the probe pulse. Specifically, the oscillation becomes slower for increasing $\Delta$ since $\Omega\propto 1/\Delta$. The initial amplitude of the forward emission also reduces over time due to spontaneous Raman decay and dephasing of the collective state. More surprisingly, we find that for all data sets the decay rates in Fig. \ref{fig:Fig2} (a, b, c) not only depend on the probe pulse lengths, but oscillate out-of-phase with the amplitude oscillations. The overall magnitude of the rates also reduces with higher $\Delta$ as expected due to the accompanying reduction in $\kappa$ and $\Gamma$. For longer pulse lengths, the decay rate approaches a constant value as the superatom reaches an equilibrium steady state because of dephasing.

\begin{figure}[b]	
\includegraphics{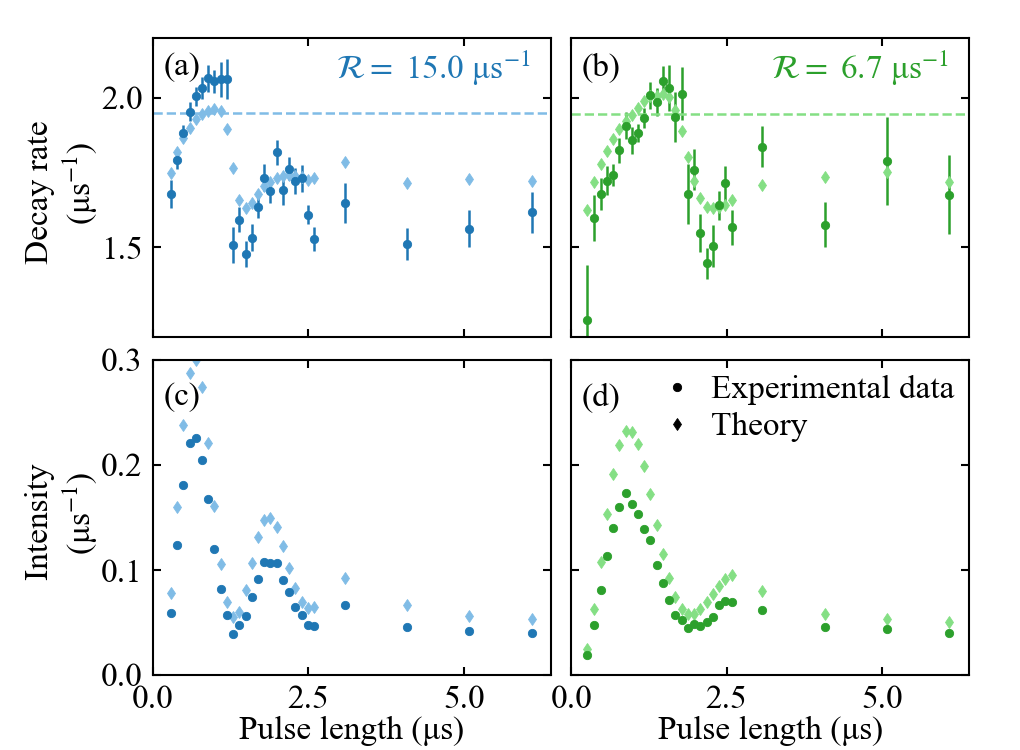}
\caption{\label{fig:Fig3} Results of fits as shown in Fig. \ref{fig:Fig1} (d) for two different photon rates with $\Delta = 2\pi\times \SI{100}{MHz}$. Panels (a,b) show the decay rates of the linear fits to experimental data (dark points) and to the model with the modified parameters as discussed in the text (light diamonds) for $\mathcal{R_{\mathrm{p}}} = \SI{15.0}{\micro s^{-1}}$ and $\SI{6.7}{\micro s^{-1}}$. The dashed lines are the results of a two-plus-one-level model as shown in Fig. \ref{fig:Fig4} (a). Panels (c,d) show the corresponding initial amplitudes of the fits to experimental data and theory. The data in (a) and (c) are the same as in Fig. \ref{fig:Fig2} (a) and (d).}
\end{figure}

Besides $\Delta$, we also investigate in Fig. \ref{fig:Fig3} the effect of changing $\Omega$ by reducing the input probe photon rate $\mathcal{R_{\mathrm{p}}}$ with $\Delta = 2 \pi \times \SI{100}{MHz}$ fixed, such that $\kappa$ and $\Gamma$ are constant.
As $\mathcal{R_{\mathrm{p}}}$ is reduced from $\SI{15.0}{\micro s^{-1}}$, Fig. \ref{fig:Fig3} (a, c), to $\SI{6.7}{\micro s^{-1}}$, Fig. \ref{fig:Fig3} (b, d), the oscillations in decay rate and initial amplitude become correspondingly slower, but the range of decay rates remains the same. For $\mathcal{R_{\mathrm{p}}}=\SI{6.7}{\micro s^{-1}}$, $\Omega$ is comparable to the case $\Delta = 2 \pi \times \SI{125}{MHz}$ and $\mathcal{R_{\mathrm{p}}}=\SI{15.0}{\micro s^{-1}}$ in Fig. \ref{fig:Fig2} (b, e) such that the oscillation periods for both values match, while the decay rate values increase as expected for stronger coupling.


The core conclusion from our observations is that the collective forward emission of the superatom depends on its internal dynamics and the single-excitation state of the ensemble at the end of the driving pulse, which in general is expected to be a superposition of the bright state $|W\rangle$ and subradiant states. Especially, we find a remarkable correlation between the decay rate and the initial intensity which translates to the probability of being in the bright state. The oscillatory behavior of the observed decay rates indicates that these internal dynamics are based on a coherent process. On a theoretical level, such a coherent term is well understood from a microscopic analysis \cite{Lehmberg1970} as the interaction between light and matter not only gives rise to the collective and directed emission, but also to a coherent interaction between the emitters. While in many situations this term can be neglected, it has recently been attracted an increased interest \cite{Chang2015b,Pichler2015,Rey2016}. Especially, it was demonstrated within a semi-classical approach that this coherent term leads to a non-exponential behavior of the emitted photons \citep{Adams2018}, as well as a remarkable universal dynamics for a one-dimensional waveguide \cite{Buchler2018} using the full quantum dynamics.

In the following, we demonstrate that an effective four-level model, shown in Fig. \ref{fig:Fig4} (a), which incorporates the dynamics of the superatom during the driving pulse, the non-trivial collective decay, as well as the intrinsic coherent coupling between the emitters and the dephasing, is capable to capture the experimentally observed phenomena. 
The core of the model is a coupling between the collective ground state $\ket{G}$ and the singly excited collective bright state $\ket{W}$ with $\Omega_{\mathrm{col}}$. The coherent coupling of the bright state to $N-1$ other singly excited subradiant eigenstates due to photon exchange as well as the dephasing of all excited states due to external decoherence are simplified in our model by introducing only two additional states. We model the exchange interactions between all excited states by including a coherent coupling with strength $\varkappa$ between $|W\rangle$ and an additional state denoted as $|C \rangle$. In the supplement, we demonstrate that for a chiral waveguide, this approximation is capable to capture with high precision the full dynamics of $N$ atoms \cite{suppMat}. Furthermore, we describe the dephasing of both the bright state $\ket{W}$ and the state $|C\rangle$ by a non-radiative decay into a dark state $|D\rangle$ with rate $\gamma_{\mathrm{D}}$. This treatment of the dephasing of the collective excited states is justified for the large number of emitters $N$ in our superatom \cite{Buchler2011,suppMat}. Finally, we include the Raman decay of all excited states to the ground state with rate $\Gamma$. 
Then, the effective model is described by a Lindblad master equation for the superatom density matrix $\rho$ is given by 
\begin{align}
\label{eq:master_equation} \partial_t \rho(t) =& - \frac{i}{\hbar} \left[ H_0(t), \rho(t) \right] + (\kappa + \Gamma) \mathcal{D}[\sigma_{\mathrm{GW}}] \rho(t) \nonumber \\
& + \gamma_{\mathrm{D}} \mathcal{D}[\sigma_{\mathrm{DW}}] \rho(t) + \Gamma \mathcal{D}[\sigma_{\mathrm{GD}}] \rho(t)\\
& + \gamma_{\mathrm{D}} \mathcal{D}[\sigma_{\mathrm{DC}}]\rho(t) + \Gamma \mathcal{D}[\sigma_{\mathrm{GC}}] \rho(t) \nonumber
\end{align}
where $\mathcal{D}[\sigma] = \sigma \rho \sigma^\dagger-(\sigma^\dagger \sigma \rho + \rho \sigma^\dagger \sigma)/2$ is the Lindblad dissipator. 
The coherent coupling is described by the Hamiltonian
\begin{align}
  H_0(t) = 2 \hbar \sqrt{\kappa\mathcal{R}_\mathrm{p}} \: \sigma_{\mathrm{GW}}^{\dag} + \hbar \varkappa \: \sigma_{\mathrm{C W}}^{\dag} + \mathrm{h.c.},
\end{align}
where $\sigma_{\mu\nu} = \ket{\mu}\bra{\nu}$. 
Note, that this four level model is a natural extension of the previous theoretical study \cite{Hofferberth2017}, and becomes necessary as we study the emitted light with much higher accuracy.

\begin{figure}	
\includegraphics{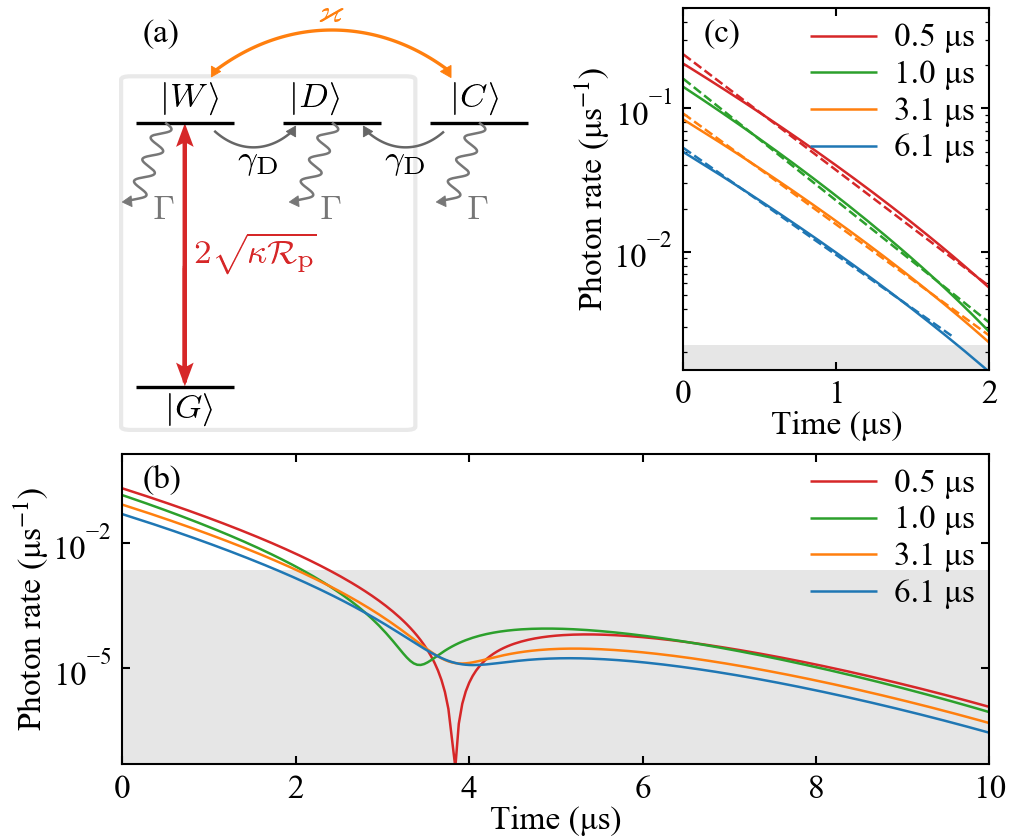}
\caption{\label{fig:Fig4} (a) Level scheme for our effective model including internal coherent dynamics of the superatom. The probe field couples the collective ground state $\ket{G}$ to the collective bright state $\ket{W}$ with an effective coupling strength $2\sqrt{\kappa \mathcal{R}_\mathrm{p}}$. The bright state $\ket{W}$ irreversibly decays with rate $\gamma_{\mathrm{D}}$ into a dark state $\ket{D}$, which does no not couple to the light. In addition, $\ket{W}$ coherently couples to a single subradiant state $\ket{C}$. This new subradiant state also irreversibly decays to $\ket{D}$. All excited states decay via Raman decay $\Gamma$ to the ground state. (b) Collective emission into the forward direction after the driving pulse has ended as predicted by this model for different pulse length calculated for the same parameters as found for the data shown in Fig. \ref{fig:Fig1} (c, d). The shaded area shows the cut-off region of Fig. \ref{fig:Fig1} (d). (c) Zoom of (b) to the experimentally accessible region shown in Fig. \ref{fig:Fig1} (d). The dashed lines show exponential fits as described in the text.}
\end{figure}

The master equation in Eq.~(\ref{eq:master_equation}) can be solved numerically and directly compared to the experimental data. Importantly, this simple model describes the superatom Rabi oscillations while the system is driven, while also predicting the time varying collective decay after the driving pulse. Specifically, the predicted decay rate after the probe pulse for the pulse lengths and parameters for the data in Fig. \ref{fig:Fig1} are shown in Fig. \ref{fig:Fig4} (b). We find that over the full decay time, the forward emission becomes clearly non-exponential and instead features a drop and revival, which is strongest for the shorter pulses; in agreement with recent predictions in \cite{Adams2018}. This results from coherent excitation shelving into $\ket{C}$ where the excitation is protected from forward emission until it is transferred back into $\ket{W}$. For longer pulses, the feature becomes less prominent as the dynamics are dominated by dephasing, which also prevents the occurrence of a second emission drop. As indicated by the gray-shaded area, the predicted dip in emission occurs at photon rates two orders of magnitude below the noise level in the experiment. Thus, we focus in Fig. \ref{fig:Fig4} (c) on the predicted emission in the experimentally observable early part of the decay. Here, the model already predicts a slight deviation from purely exponential decay, but comparing to the data in Fig. \ref{fig:Fig1} (d), this tendency in the slope of the emission is not visible with our current experimental resolution. To enable comparison between the model and experiment, we extract a decay rate and initial intensity from the numerical solutions in the same way as for the experimental data. The results are shown by the dashed lines in Fig. \ref{fig:Fig4} (c). 

With this approach, we obtain the model predictions for initial slope and intensity of the superatom decay shown in Figs. \ref{fig:Fig2} and \ref{fig:Fig3}. In our optimization procedure, we assume that $\gamma_\mathrm{D}$ is the same in all datasets and that $\kappa$ scales with the single-photon detuning as $\kappa \propto 1/\Delta^2$. The Raman decay rate $\Gamma$ is determined from experimental parameters. The obtained fitting parameters for $\kappa$, $\varkappa$ and $\gamma_\mathrm{D}$ are given in \cite{suppMat}. We find that our simple model captures the oscillatory behaviour of the decay rate as a function of pulse length quite well for all datasets. Furthermore, the values found for the coherent coupling $\varkappa$ are of the same order as the collective coupling $\kappa$, as expected from the comparison of our model to microscopic simulations of an idealized one-dimensional system \cite{suppMat}. 

We also show the constant decay rates predicted by the model if we disable the coherent coupling by forcing $\varkappa=0$ (dashed lines in Fig. \ref{fig:Fig2}), to show that the coherent coupling is the crucial factor to explain the experimental observations. In previous work, we successfully described single-photon absorption \cite{Buchler2011,Hofferberth2016e} and photon correlations mediated by the superatom \cite{Hofferberth2017,Hofferberth2018a} without the coherent coupling. 
These observations are still captured by the 4-level model presented here. However, the influence of the coherent population shelving only becomes visible in the precise study of the decay dynamics after the driving pulse.

In summary, we observed non-trivial collective emission dynamics of a Rydberg superatom, which we attribute to coherent exchange interactions between individual atoms. We show that a simple effective model including a single coherently coupled subradiant state reproduces the experimental observations quite well. Our work is complementary to recent investigations of weakly excited ensembles \cite{Pillet2018,Adams2018,Adams2018b} and structured emitter arrays \cite{Imamoglu2018,Park2018,Bloch2020}, but adds a new component through the saturation of the ensemble by a single photon because of the Rydberg blockade, which imposes further challenges for a full theoretical treatment. 
While our model is motivated by comparison to simulations of an idealized waveguide system and captures the core aspects of the system, a full understanding of the coherent and incoherent dynamics in a thermal atomic ensemble will ultimately require a full microscopic model. Our observations are of immediate consequence for the study and application of Rydberg superatoms and other collective quantum emitters, for example in cascaded emitter systems in waveguide-like geometries, where the internal dynamics will significantly alter the behaviour of the full ensemble \cite{Buchler2018}. More generally, these dynamics become relevant whenever collective excitations are created or probed on timescales comparable to the coherent photon exchange rates, e.g. in quantum simulation or photon memories. On one hand, it will be relevant to study to what extend such internal interaction dynamics impose a fundamental limit on applications of collective excitations for single photon sources and quantum gates. At the same time, a better understanding of these dynamics enable precise collective state engineering, for example to efficiently store photons in sub-radiant collective states \cite{Chang2018}. 

\textit{Acknowledgements} We thank Charles Adams, Robert Bettles, Thomas Stolz, Thomas Pohl, and Klaus Mølmer for fruitful discussion. This work is supported by the European Union's Horizon 2020 program under the ERC consolidator grants SIRPOL (grant N. 681208) and RYD-QNLO (grant N. 771417), the ErBeStA project (No. 800942), grant agreement No. 845218 (Marie Sk\l{}odowska-Curie Individual Fellowship to H.B.), and the Deutsche Forschungsgemeinschaft (DFG) under SPP 1929 GiRyd project BU 2247/4-1.

\bibliography{single_superatom_afterlight}

\clearpage
\onecolumngrid

\begin{center}
    \Large{Supplementary Information: \\Observation of collective decay dynamics of a single Rydberg superatom}
\end{center}

\maketitle

\setcounter{figure}{0} \renewcommand{\thefigure}{S.\arabic{figure}} 
\section{Experimental details}
Below, we give experimental details in addition to the main text, summarizing the ensemble preparation and the experimental sequence that we employ subsequently.

\subsection{Ensemble preparation}
We start from a cigar shaped ensemble of $^{87}$Rb atoms in a crossed optical dipole (wavelength $1070\,\mathrm{nm}$, $1/e^2$-waist $\approx 55\,\mathrm{\mu m}$, intersection angle $30^{\circ}$) loaded from a magneto-optical trap (MOT).
Following a final compression of the MOT, the atoms are evaporatively cooled as we reduce the trap light intensity in two stages.
For additional cooling and to reduce atom loss, we employ Raman sideband cooling for $16\,\mathrm{ms}$ during each of the linear evaporation ramps.
Longitudinal confinement along the probe axis below the range of Rydberg blockade is provided by a tightly focused optical trap with an elliptical cross-section (wavelength $805\,\mathrm{nm}$, $1/e^2$-waists $\approx 10\,\mathrm{\mu m}$ and $\approx 21\,\mathrm{\mu m}$) that intersects perpendicularly with the cigar-shaped ensemble as well as the probe and control beams at their focus. In the dimple trap, the atomic temperature is $\approx \SI{10}{\micro K}$.
The dimple trap intensity is kept constant throughout the evaporation process and the dimple position can be fined tuned with respect to the probe using an acousto-optical deflector.
Before starting experiments, the crossed dipole trap intensity is ramped to zero such that atoms outside the dimple are released and move outside the experimental region before increasing it again to provide confinement in the radial probe direction for the atoms inside the dimple.
In combination with the $1/e^2$-waist radius of the probe ($\approx 6.5\,\mathrm{\mu m}$), the dimple confinement restricts the excitation volume below the blockade range.
The focus of the control beam is larger ($\approx 14\,\mathrm{\mu m}$) to limit variation of $\Omega_c$ across the excitation volume.

\subsection{Superatom excitation and probe photon detection}
\begin{figure}
\includegraphics{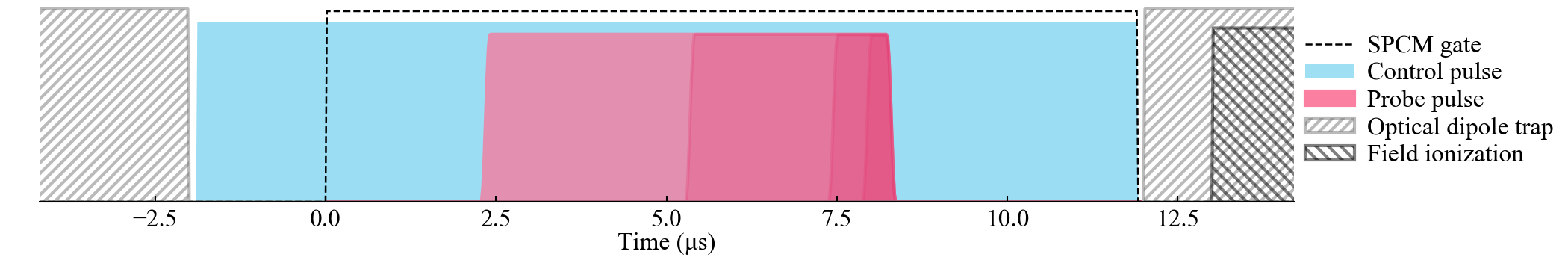}
\caption{\label{fig:Fig5} Sketch of experimental procedure. The crossed optical dipole trap is turned off for $\SI{14}{\micro s}$. As the dipole trap turns off, the Rydberg control laser is turned on. We employ a $\SI{2}{\micro s}$ wait time before the gates of the single photon counter modules are opened to ensure that the control light is fully on. Probing is done during the $\SI{12}{\micro s}$ where the single photon counter gates are open. We keep the end-time of the probe pulse fixed and vary the point in time when the probe pulse is turned on. The dipole trap is turned back on as of the gates are turned off, and a subsequent field ionization pulse is applied to remove any remaining Rydberg excitation in the system. Produced ions are detected by a multi-channel plate.}
\end{figure}

Following preparation of the atomic ensembles in the dimple, we employ the experimental sequence shown in Fig \ref{fig:Fig5}.
The crossed dipole trap is turned off every $100\,\mathrm{\mu s}$ for $14\,\mathrm{\mu s}$. 
The dimple potential remains present throughout the experiment to restrict motion along the axial probe direction.
We account for the resulting AC-Stark shift by adjusting the the probe frequency accordingly.
The shift also suppresses Rydberg excitation of atoms outside the dimple following release from the crossed-dipole trap.

The control field is turned on as the dipole trap is turned off, and after a $\SI{2}{\micro s}$ wait time the gates of the SPCMs are opened and probing takes place. In the experiments described here, we keep the probe pulse end-time fixed and vary the start-time.

Following conclusion of a single experimental shot, an electric field pulse ionizes any remaining Rydberg atoms to avoid the presence of residual Rydberg excitations during the next iteration of the superatom excitation. The ions are detected on a multi-channel plate.
Overall, we conduct $1000$ individual experimental cycles. Afterwards the atoms are leased from the trapping potentials by turning the optical dipole trap and the dimple trap off for $\SI{10}{m s}$. The traps are then turned back on and $1000$ probing cycles without atoms present are performed to acquire reference pulses of the probe. 
Subsequently, we prepare a new atomic ensemble repeating the procedure above.

The light emitted by the superatom is detected using four single photon counting modules (SPCMs) in a Hanbury-Brown Twiss configuration located behind a single-mode optical fibre, which is aligned onto the original probe mode and acts as a spatial mode filter.
The overall detection efficiency including the SPCM quantum efficiency and all optical loss between the experimental region and the SPCMs is $\approx 35\% $.

For each of the four datasets presented in the main text, we take data for 26 different pulse lengths. For the dataset with $\Delta = 2\pi \times \SI{100}{MHz}, \mathcal{R}_{\mathrm{p}} = \SI{15.0}{photons/\micro s}$, we take $1 111 \times 10^3$ measurements for each pulse length. For $\Delta = 2\pi \times \SI{125}{MHz}, \mathcal{R}_{\mathrm{p}} = \SI{15.0}{photons/\micro s}$, we take $621\times 10^3$ measurements, and for $\Delta = 2\pi \times \SI{150}{MHz}, \mathcal{R}_{\mathrm{p}} = \SI{15.0}{photons/\micro s}$ we take $467\times 10^3$ measurements. For $\Delta = 2\pi \times \SI{100}{MHz}, \mathcal{R}_{\mathrm{p}} = \SI{6.7}{photons/\micro s}$, we take $377\times 10^3$ measurements.

In the main text, we discuss a threshold on the values included in the fits shown in the main paper in Fig. 1 (d). The threshold is the point where the statistical uncertainty on the datapoints become similar to their values. The threshold is in this case defined as the value where less than 50 counts has been detected in one time-bin of 20 ns in all the measurements.

\section{Superatom model with internal coherent dynamics}
In the main text we introduce an effective model with four effective states, see Fig. 4, which attempts to capture both coherent and incoherent internal dynamics of the superatom in the simplest possible way. The main feature of this model is a coherent coupling between a collectively excited bright state and a single collectively excited subradiant state to represent the coherent dynamics inside the superatom, while a second effectively dark state represents weakly coupled collective states into which the system can dephase. We show in the following, how this effective model is motivated by the microscopic dynamics of a system of $N$ emitters coupled to a single-mode light field.

Generally, only one excited atomic state $|\mathrm{W}\rangle$ is coupled to the specific mode of the light field and emits collectively back into the same mode. However, $|\mathrm{W}\rangle$ couples coherently to $N-1$ subradiant states $|\mathrm{C}_i \rangle$ via exchange of virtual photons inside the excited ensemble. The strength of this coherent coupling is different between the different eigenmodes of the system, and depends on the individual atom positions. To study the effect of the coherent exchange, we consider a system of $N$ quantum emitters coupled to a one-dimensional chiral waveguide. The chiral one-dimensional waveguide has many similarities to the Rydberg superatom \cite{Buchler2018} and can be solved analytically in the case of stationary emitters. 

In the one-dimensional, chiral waveguide the atoms are coherently coupled to each other by the one-dimenisonal dipole-dipole interaction
\begin{equation}
    H_{\mathrm{exc}} =
        \frac{i\hbar \kappa}{2 N} \sum_{l,j} \mathrm{sign}(k_0(x_l - x_j)) \sigma_{l,j},
\end{equation}
which describes the exchange of a virtual photon from atom $j$ to atom $l$, mediated by the operator $\sigma_{l,j}$. As stated above, the bright state $|\mathrm{W}\rangle = \sum |j\rangle/ \sqrt{N}$, where $\ket{j}$ indicates that the $j$'th atom is in the Rydberg state, is distinct from all other excited states, in that it is the only state coupled to the light field. Therefore, it is practical to re-express the exchange Hamiltonian in a basis containing the bright state and orthogonal subradiant states. After diagonalization of the subradiant state subspace, the exchange Hamiltonian in this basis becomes
\begin{equation}
    H_{\mathrm{exc}} =
        \hbar\sum_{j=1}^{N-1} \big(\kappa_j\ \sigma_{\mathrm{W},\mathrm{C}_j} + \mathrm{H.c.}\big)
        + \hbar\sum_{j=1}^{N-1} \epsilon_j\ \sigma_{\mathrm{C}_j,\mathrm{G}} \sigma_{\mathrm{G},\mathrm{C}_j},
\end{equation}
where $\kappa_j = \kappa[i + \cot(\pi j/N) ]/ 2N$ and $\epsilon_j = - \kappa\cot(\pi j/N)/(2N)$ are coupling strengths between atomic states. $|\mathrm{C}_j\rangle$ is one of the $N-1$ subradiant states. In this basis we directly see that there are subradiant states with weak coupling to the bright state of order $\mathcal{O}(1/N)$. Another contribution to the dynamics is the driving of the atomic system with incident light with amplitude $\alpha(t)$, which couples the ground state to the bright state
\begin{equation}
    H_{\mathrm{drive}} = \hbar\sqrt{\kappa} \alpha(t) \big(\sigma_{\mathrm{W},\mathrm{G}}
        + \sigma_{\mathrm{G},\mathrm{W}}\big).
\end{equation}
Here, $\kappa$ is the bright state enhanced coupling to the driving light field. The excited system will eventually decay. However, only the bright state will decay back into the waveguide. Coming from the full microscopic theory, this decay rate is given by $\kappa$. Altogether, the dynamic of the chiral waveguide model is described by the master equation
\begin{equation}
    \partial_t \rho = - \frac{i}{\hbar} [H_{\mathrm{exc}}+H_{\mathrm{drive}}, \rho]
        + \kappa \mathcal{D}_{\sigma_{\mathrm{G},\mathrm{W}}} [\rho].
    \label{eq:masterEqChiral}
\end{equation}

\begin{figure}[t]
\includegraphics{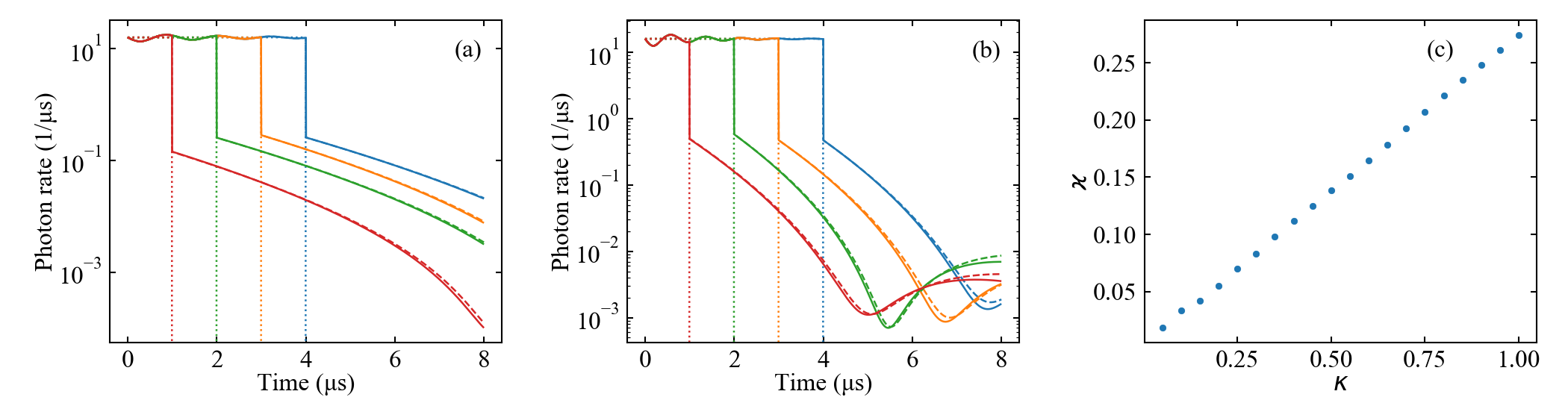}
\caption{\label{fig:Fig6} (a) and (b) Emission of light from $N=1000$ atoms coupled to a one-dimensional waveguide for four pulse lengths and with (a) $\kappa = \SI{0.45}{\per\micro s}$ and (b) $\kappa = \SI{1.0}{\per\micro s}$. Solid lines are results from the waveguide, dashed lines are fits with the four-level model to the waveguide results, and dotted lines show where the driving field ends. $\Gamma = \SI{0.1}{\per\micro s}$. Comparing to experimental data, the value of $\kappa$ in (a) is the same as we find in the experiment, whereas (b) shows a higher $\kappa$ but $\varkappa$ similar to what is found in experiment.
(c) By fitting the four-level model to the chiral waveguide data for different values of $\kappa$ it is possible to extract a scaling of $\varkappa$ with $\kappa$.} 
\end{figure}

We can further account for spontaneous decay of single atoms in other directions than the waveguide with a constant rate $\Gamma$. In the bright state-subradiant state basis this adds a decay term $\Gamma \mathcal{D}_{\sigma_{\mathrm{G},\mathrm{C_i/W}}}[\rho]$ for each subradiant state and the bright state. 

Solving the master equation of the chiral waveguide model~\eqref{eq:masterEqChiral}, we can calculate the light emission from $N=1000$ emitters for varying pulse lengths and for different $\kappa$, with examples shown by the solid lines in Fig. \ref{fig:Fig6} (a) and (b). We find that these calculations indeed show varying forward-directed decay of the ensemble as observed in the experiment.

Next, we can test how well our simplified effective model with only a single coherently coupled state reproduces these results. Towards this end, we fit the numerical calculations with our effective model in the same way as we fit the experimental data, with $\Gamma$ fixed to the calculated Raman lifetime to obtain the model parameters $\kappa$ and $\varkappa$. The best fits of the effective model are shown as dashed lines in Fig. \ref{fig:Fig6} (a) and (b). We find that reducing the $N-1$ collective subradiant states to a single coherently coupled level works extremely well for the chiral waveguide case. We can understand this from the above observation that only a few subradiant states have sufficiently strong coupling to the bright state to contribute significantly to the coherent part of the dynamics on the timescale of the collective decay. Moreover, from analyzing simulations with different collective coupling strength to the waveguide, we find a linear scaling between $\kappa$ and $\varkappa$ for the best fits of our model, as shown in Fig. \ref{fig:Fig6} (c). 

The considered system of stationary emitters in the chiral one-dimensional waveguide theory contains no sources of dephasing or decoherence except the single-atom spontaneous decay. In the experiment, multiple sources contribute to finite decoherence on the relevant time scale. First, we assume thermal motion to play a crucial role. For our ensemble temperature $T = \SI{10\pm 1}{\micro K}$, we estimate that the atoms have a most likely velocity of $v \approx 0.035 \lambda/\SI{}{\micro s}$, where $\lambda =2 \pi \left(\frac{1}{k_\mathrm{p}} - \frac{1}{k_\mathrm{c}} \right)$ is the spin wavelength imprinted by the light into the ensemble. $k_\mathrm{p}$ and $k_\mathrm{c}$ are the wave vectors of the probe and the control fields respectively. The sign difference arises because the two fields are counter-propagating.
This means that during one measurement, the atoms move about half of a spin-wavelength. Adding thermal motion to our one-dimensional model in the form of Boltzmann distributed velocity for each atom does not alter the dynamics significantly. This is specific for the chiral waveguide, since here the time-dependent phase each atom picks up due to its velocity can be transformed into a detuning, effectively doubling the Doppler shift. This is not possible in a non-chiral or three-dimensional system. With this in mind, thermal motion may be added by an effective dephasing model, mixing the excited states with each other. Other sources of dephasing in the experiment can originate from finite laser linewidth or fluctuating electric fields in the experiment chamber.

We thus need to include additional dephasing in our effective model and the chiral waveguide model to reproduce the experimental results. Dephasing effectively shifts population from the bright state and the few strongly coupled subradiant states into the many weakly coupled subradiant states. For large atom number $N$, this can be treated as an effective decay into the uncoupled states \cite{Buchler2011} as any form of revival will not happen on the experiment timescale \cite{Hofferberth2016e}. We implement this in our effective model by adding an additional dark state that acts as a reservoir into which both bright and subradiant state can decay with rate $\gamma_\mathrm{D}$.

\section{Experiment parameter estimation}
Here, we outline the procedure used to obtain the parameters used for the results of the effective four-level model presented Fig.s 2 and 3. First, we fix $\Gamma = \Gamma_{\mathrm{e}} (\Omega_{\mathrm{c}}/2\Delta)^2$ based on the known values for $\Delta$ and $\Omega_{\mathrm{c}}=2\pi\times\SI{13}{MHz}$. We then fit the four-level model with $\varkappa = 0$ to the the full probe pulse as in previous works \cite{Hofferberth2017} to extract $\kappa$ and verify that $\kappa\sim 1/\Delta^2$ for the different datasets.
To determine $\gamma_{\mathrm{D}}$ and $\varkappa$, the full four-level model shown in Fig. 4 (a) is fitted to the data. 
Since the fits put more weight on the points within the pulses, the values found for $\gamma_{\mathrm{D}}$ tend to be overestimated. We attribute this overestimation to additional decoherence channels which the model does not account for, such as laser linewidth.
Therefore $\gamma_{\mathrm{D}}$ is adjusted separately to reproduce the observed emission rate after the probe pulse is over.
We find a common value for all datasets since dephasing is dominanted by atomic motion, which is independent of $\mathcal{R}_{\mathrm{p}}$ and $\Delta$.
Table \ref{tab:tab2} lists the full parameter sets for the theory data in Fig.s 2 and 3. 

\begin{table}[ht]
\caption{\label{tab:tab2}Parameters sets for theory data. }
\begin{tabular*}{0.8\textwidth}{r @{\extracolsep{\fill}} rrrrrc}
$\mathcal{R}$ $(\SI{}{\micro s^{-1}})$  & $\Delta/2\pi$ $(\SI{}{\micro s^{-1}})$ & $\kappa$ $(\SI{}{\micro s^{-1}})$  & $\Gamma$ $(\SI{}{\micro s^{-1}})$  & $\gamma_\mathrm{D}$  $(\SI{}{\micro s^{-1}})$   & $\varkappa$  $(\SI{}{\micro s^{-1}})$  & In figures \\
\hline
15.0 & $ 100$   & $0.46$   & $0.15$     & $0.85$    & $0.31$    & 1,2,3,4\\
15.0 & $ 125$   & $0.32$   & $0.10$     & $0.85$    & $0.32$    & 2\\
15.0 & $ 150$   & $0.21$   & $0.064$    & $0.85$    & $0.31$    & 2\\
6.7 & $ 100$    & $0.47$   & $0.15$     & $0.85$    & $0.34$    & 3\\
\end{tabular*}
\end{table}

Interestingly, we find approximately the same value of $\varkappa$ for the three different values of $\kappa$ discussed in this work. We do not find a scaling of $\varkappa$ with $\kappa$ as it is the case for the previously discussed one dimensional chiral waveguide model. 

\twocolumngrid


\end{document}